\documentstyle[12pt]{article}
\begin{document}
\noindent
\begin{center}
{\Large {\bf Hawking Radiation in Trace Anomaly Free Frames }}

\vspace{2cm} {\bf H.~
Salehi}\footnote{e-mail:~h-salehi@sbu.ac.ir.} \vspace{0.5cm}and
\vspace{0.5cm}{\bf S.~ Mirabotalebi}\footnote{e-mail:~s-mabotalebi@sbu.ac.ir.} \vspace{0.5cm}\\
{\small {Department of Physics, Shahid Beheshti University, Evin,
Tehran 19839,  Iran.}}\\
\end{center}
\vspace{1cm}

\begin{abstract}

We have used the results of renormalization of a two-dimensional
quantum stress tensor to develop a conformally invariant dynamical
model. The model requires the consideration of those conformal
frames in which there exists a correspondence between the trace
anomaly and a cosmological constant. We apply this model to a two
dimensional-Schwarzschild (-de Sitter) spacetime to show that in
these conformal frames one may achieve Hawking radiation without
recourse to the trace anomaly .
\end{abstract}


\section{Introduction}
In the semiclassical approximation of quantum gravity \cite{1} the
expectation value of a quantum stress-tensor is used as the source
of the Einstein field equations. In order to have a well defined
stress-tensor, renormalization must be carried out due to
intrinsically divergent nature of products of field operators at
the same spacetime point. The remarkable consequences of all the
renormalization programs which have been developed so far is that
the stress-tensor of a conformally invariant field obtains a
nonvanishing trace \cite{2}.There have been many attempts to
understand the physical implications of such an anomalous trace.
Specifically, it has been shown \cite{3} that the imposition of a
regularity condition of a covariantly conserved stress tensor on
the horizon of a Schwarzschild black hole would lead to a
connection between the trace anomaly and Hawking radiation in two
dimensions. In view of this observation one may attribute a
nonlocal characteristic to the trace anomaly in the sense that  it
relates the properties of the quantum stress tensor on the horizon
to those all those at null infinity. The nonlocal characteristic
of the trace anomaly has been already investigated in some
different context \cite{5,6}. Here we wish to study this property
by introducing a two-dimensional conformally invariant dynamical
model. The use of this model in the two-dimensional Schwarzschild
(-de Sitter) spacetime reveals that the existence of the trace
anomaly may not be a necessary condition for obtaining the Hawking
effect.

We have organized this paper as follows: In section $2$, we use
the results of renormalization of a quantum stress-tensor in
two-dimensions to develop a conformally invariant dynamical model.
The conformal invariance allows us to consider the model in
different conformal frames. In particular, we shall obtain a
consistency requirement enforcing the consideration of those
frames in which the trace anomaly takes the constant value
$\Lambda$. In section $3$, we first apply this model to a two
dimensional Schwarzschild spacetime by considering a class of
conformal frames characterized by $\Lambda=0$. By exploring the
conservation law of the quantum stress-tensor in these frames, in
which the trace anomaly vanishes, we obtain an outward flux of
thermal radiation with temperature $kT_{H}=(8\pi M)^{-1}$. We then
bring the case $\Lambda\neq 0$. We shall show that in the
corresponding conformal frames the Hawking temperature receives a
correction term as a result of the existence of $\Lambda$. In
section 4, we offer some concluding remarks. We shall work in
units in which $c=\hbar=1$, and our sign conventions are those of
Hawking and Ellis \cite{7}.


\section{The model}

We consider a massless quantum  scalar field conformally coupled
to a two-dimensional gravitational background. Renormalization of
the corresponding stress-tensor $\Sigma_{\mu\nu}$ leads to the
following results \cite{8}
\begin{equation}
\nabla^{\mu}\Sigma_{\mu\nu}=0 \label{a1}
\end{equation}

\begin{equation}
\Sigma_{\mu}^{\mu}=\frac{1}{24\pi} R \label{a2}\end{equation}
where $\nabla$ denotes the covariant differentiation  operator,
and $R$ is the curvature scalar. The first equation indicates that
$\Sigma _{\mu\nu}$ is covariantly conserved. The second one is the
trace anomaly coming from the renormalization process, and suggest
that $\Sigma_{\mu\nu}$ can be written in the form
\begin{equation}
\Sigma_{\mu\nu} = \Sigma_{\mu\nu}^{(0)}+\frac{1}{48 \pi}
g_{\mu\nu} R \label{a3}\end{equation} where
$\Sigma_{\mu\nu}^{(0)}$ is a traceless tensor. The plane is to
construct a dynamical model based on the decomposition (\ref{a3}),
in the close correspondence to \cite{5,6}. The essential step is
to identify $\Sigma_{\mu\nu}^{0}$ with the stress-tensor of a
conformally invariant scalar field

\begin{equation}
T_{\mu\nu}[\phi]= \nabla_{\mu}\phi\nabla_{\nu}\phi -
\frac{1}{2}g_{\mu\nu}\nabla_{\gamma}\phi\nabla^{\gamma}\phi
\label{a5}\end{equation} where $\phi$ is a C-number scalar field,
satisfying
\begin{equation}
\Box\phi=0 \label{f}\,.\end{equation} The traceless condition of
$T_{\mu\nu}$ is satisfied. The relation (\ref{a3}) then leads
\begin{equation}
\Sigma_{\mu\nu}=T_{\mu\nu}+\frac{1}{48\pi}g_{\mu\nu} R
\label{a6}\end{equation} This may now interpreted as a general
condition imposed on $\Sigma_{\mu\nu}$ for a fixed background
geometry. Nevertheless, from comparing Eq.(\ref{a1}) with
Eq.(\ref{a6}) one infers that such an interpretation is not
consistent on the background metric, for the tensor $T_{\mu\nu}$
can't be conserved and receives a source term given by the trace
anomaly. This discrepancy may be removed by appealing to the
conformal invariance of the above model, implying that the
coupling of $\phi$ to geometry can be studied in many different
conformal frames which are dynamically equivalent. A generic
question which appears in these situation is that which of these
frames corresponds to the physical one. In the present case we
wish to find a particular conformal frame in which $T_{\mu\nu}$
can be expressed as a conserved stress-tensor. To this end we
first consider a conformal transformation
\begin{equation}
\bar{g}_{\mu\nu}=\Omega^{2}(x) g_{\mu\nu} \label{a7}\end{equation}
\begin{equation}
\bar{\phi}(x)\longrightarrow\Omega^{-1}(x)\phi(x)
\label{a8}\end{equation} under which (\ref{a7}) takes the form

\begin{equation}
\bar{\Sigma}_{\mu\nu}=\bar{T}_{\mu\nu}+ \frac{1}{48
\pi}\bar{R}\bar{g}_{\mu\nu} \label{a9}\,.\end{equation}

We then restrict our attention to those frames in which the trace
anomaly can be related to some constant value. These frames are
defined by

\begin{equation}
\bar{R}=\Lambda \label{b2}\end{equation} or equivalently,

\begin{equation}
\Omega^{-2}\{R+2\nabla_{\gamma}\ln\Omega\nabla^{\gamma}\ln\Omega
-2\frac{\Box\Omega}{\Omega}\}=\Lambda \label{b3}\end{equation}
where we take $\Lambda$ to be a cosmological constant. This
relation is actually a constraint on the conformal factor. It
determines the class of conformal frames in which
$T_{\mu\nu}$appears as a conserved tensor. This can be seen by
combining the equations(\ref{a9},\ref{b2})which gives
\begin{equation}
\bar{\Sigma}_{\mu\nu}=\bar{T}_{\mu\nu}+\frac{1}{48\pi}\Lambda\bar{g}_{\mu\nu}.
\label{b4}\end{equation}

Our next step is to study the relation (\ref{b4}) in a particular
case corresponding to a two dimensional Schwarzschild(de-Sitter)
spacetime.

\section{Hawking Effect}

Consider first the case $\Lambda=0$, we intend to apply the above
model to a two-dimensional Schwarzschild spacetime, described by
the metric
\begin{equation}
ds^{2}=-(1-\frac{2GM}{r})dt^{2}+(1-\frac{2GM}{r})^{-1}dr^{2}
\label{b5}\end{equation} where $M$ is the mass of the black hole
and $G$ is the gravitational constant. This metric can be written
in the conformally flat form
\begin{equation}
ds^{2}=h(r)(-dt^{2}+dr^{* 2}) \label{b6}\end{equation} where :
\begin{equation}
h(r)=(1-\frac{2GM}{r})  \,\,,\,\,  \frac{dr}{dr^{*}}=h(r)
\label{b7}\end{equation} A conformal transformation transforms
(\ref{b6}) to

\begin{equation}
\bar{ds}^{2}=\bar{\Omega}^{2}(r)(-dt^{2}+dr^{* 2})  \,\,,\,\,
\bar{\Omega}^{2}(r)=h(r)\Omega^{2}(r)\label{q}\end{equation} The
corresponding conformal factor can be determined by substituting
$R=\frac{4GM}{r^{3}}$ and $\Lambda=0$ into the
constraint(\ref{b3}) which leads to
\begin{equation}
\Omega^{-2}(r)\{
\nabla_{\gamma}\ln\Omega(r)\nabla^{\gamma}\ln\Omega(r)
+\frac{2GM}{r^{3}}-\frac{\Box\Omega(r)}{\Omega(r)}\}=0
\label{b8}\end{equation} This is now a differential equation for
determining the function $\Omega(r)$. We subject the solutions of
(\ref{b8})to the condition that the geometry of the quasi-flat
regions $(r_{qf}=r\gg 2GM)$ coincides with that of the background
frame, namely $\Omega(r)\rightarrow 1$ when $r\rightarrow r_{qf}$.
On these regions we must have $\bar\Omega(r)\rightarrow 1$. We may
obtain the general form of $\bar{\Sigma_{/mu/nu}}$ by using the
results of the appendix. This gives
\begin{equation}
\bar{\Sigma}_{\mu}^{\nu}=\bar{\Sigma}_{\mu}^{(r)\nu}+\bar{\Sigma}_{\mu}^{(eq)\nu}.
\label{b9}\end{equation}The tensors $\bar{\Sigma}_{\mu}^{(r)\nu}$
and $\Sigma_{\mu}^{(eq)\nu}$ are given by

\begin{equation}\label{r}
 \bar{\Sigma}^{(r)\nu}_{\mu} = K \bar{\Omega}^{-2}(r)\left(\begin{array}{cc}
-1  &  -1 \\
 1  &  1
\end{array}\right) \,,\,
\Sigma^{(eq)\nu}_{\mu}= Q \bar{\Omega}^{-2}(r)
\left(\begin{array}{cc}
-1  &  0 \\
 0  &  1
\end{array}\right)\label{r1}
\end{equation}
where $K$ and $Q$ are constants. the tensor
$\Sigma_{\mu}^{(eq)\nu}(r\rightarrow r_{qf})$ is proportional to
the stress-tensor of a gas under equilibrium condition
\begin{equation}\frac{\pi}{6} (k T)^{2}\left(\begin{array}{cc}
-1  &  0 \\
 0  &  1
\end{array}\right).\label{r2}\end{equation}
It indicates a thermal bath arising from the existence of a
cosmological event horizon \cite{9,10}.In the present case, since
$\Lambda =0$ we should have $Q=0$, requiring a vanishing
temperature. On the other hand, the tensor
$\bar{\Sigma}_{\mu}^{(r)\nu}(r\rightarrow r_{qf})$ contains off
diagonal (flux) components and should be compared with the
stress-tensor of thermal radiation
\begin{equation}\frac{\pi}{12}(k T)^{2}\left(\begin{array}{cc}
-1  &  -1 \\
 1  &  1
\end{array}\right)\label{r3}\end{equation}
It gives a temperature $k T_{H}=(8\pi M)^{-1}$ if $K=(768
M^{2})^{-1}$. It means that in the conformal frame which is
characterized by anomaly cancellation, one obtains an outward flux
of radiation with the temperature consist with the Hawking
radiation.

We now consider the case $\Lambda\neq 0$. In this case the back
ground metric should be considered to be the Schwarzschild-de
Sitter spacetime. Thus the conformal factor of the metric
(\ref{b5}) should be taken to be
\begin{equation}
h(r)=\left(1-\frac{2GM}{r}-\frac{\Lambda r^{2}}{3}\right)
\label{r4}\end{equation} A conformal transformation transforms
(\ref{b6}) to the form (\ref{q}). Using the results of the
Appendix, the general form of $\bar{\Sigma}_{\mu\nu}$ in this case
may be written as
\begin{equation}\Sigma^{(r)\nu}_{\mu} = K \bar{\Omega}^{-2}(r)\left(\begin{array}{cc}
-1  &  -1 \\
 1  &  1
\end{array}\right)+ \frac{\Lambda}{48\pi}\left(\begin{array}{cc}
1+\bar{\Omega}^{2}(L)\bar{\Omega}^{-2}(r)  &  0 \\
 0  &  1-\bar{\Omega}^{2}(L)\bar{\Omega}^{-2}(r)
\end{array}\right)\label{r5}\end{equation}

\begin{equation} \bar{\Sigma}^{(eq)\nu}_{\mu} = Q \bar{\Omega}^{-2}(r)\left(\begin{array}{cc}
-1  &  0 \\
 0  &  1
\end{array}\right)\label{r6}\end{equation}
where we have used
\begin{equation}
H(r)=\frac{\Lambda}{48\pi}\left(
\bar{\Omega}^{2}(r)-\bar{\Omega}^{2}(L)\right).\label{r7}\end{equation}
We define the quasi-flat regions as
\begin{equation}
2GM\,\ll\,r_{qf}\,\ll\,\frac{1}{\sqrt{\Lambda}}.
\label{r8}\end{equation} For a sufficiently small $\Lambda$, one
may define a domain of many orders of magnitude on which
(\ref{r8})is valid. We may impose $\bar{\Omega}(r)\rightarrow 1$
when $r\rightarrow r_{qf}$ as the boundary condition. The tensor
$\bar{\Sigma}_{\mu}^{(eq)\nu}(r\rightarrow r_{qf})$ describes a
thermal bath coming from the cosmological event horizon. Thus
comparing it with (\ref{r2}) gives a temperature $k
T_{c}=\sqrt{\Lambda}$ if we choose $Q=\frac{\pi}{6}\Lambda$. The
tensor $\bar{\Sigma}_{\mu}^{(r)\nu}$ yields
\begin{equation}\Sigma^{(r)\nu}_{\mu}(r\rightarrow r_{qf}) =
\frac{\pi}{12} (8\pi M)^{-2}\left(\begin{array}{cc}
-1  &  -1 \\
 1  &  1
\end{array}\right)+ O(\Lambda)\label{w1}\end{equation}
if $K=(768\pi M^{2})^{-1}$, and describes an outward radiation
with temperature
\begin{equation}
k T_{H}=(8\pi M)^{-1} +O(\Lambda) \label{w2}\end{equation} The
second term on the right hand side is a correction term to the
Hawking temperature, arising from the cosmological event horizon.


\section{CONCLUDING REMARKS}

A conformally invariant dynamical model is developed, using the
standard results of renormalization of a quantum stress tensor in
two dimensions. We found it necessary to consider the model in
conformal frames in which either the trace anomaly vanishes or
takes a constant value $\Lambda$, which is regarded to be a
cosmological constant. There is a correspondence between the
conformal frames  and quasi-flat regions of a Schwarzschild and a
Schwarzschild-de Sitter spacetime respectively. We have shown that
in the conformal frames in which an anomaly cancellation takes
place we obtain an outward flux of thermal radiation with  the
Hawking temperature. This observation emphasizes  the dispensable
role of the trace anomaly for obtaining the Hawking radiation in
this frame. On the other hand, in a conformal frame in which the
trace anomaly takes a constant value we have received a correction
term to the Hawking temperature due to the presence of a
cosmological event horizon.

\section{ACKNOWLEDGMENTS}

One of us (S.M) wishes to thank Y.Bisabr for helpful discussions

\section{APPENDIX}

Here we are looking for the general form of
$\bar{\Sigma}_{\mu\nu}$ which satisfies the conservation equation

\begin{equation}\label{n1}
\bar{\nabla}^{\mu}\bar{\Sigma}_{\mu\nu}=0
\end{equation}
for the metric (\ref{b7}). We first note that the nonzero
Christoffel symbols of (\ref{b7}) are

\begin{equation}
\Gamma^{r^{*}}_{tt}=\Gamma^{t}_{tr^{*}}=\Gamma^{t}_{r^{*}t}=\Gamma^{r^{*}}_{r^{*}r^{*}}
=\frac{1}{2}\frac{d}{dr}\bar{\Omega}(r) \label{n2}\end{equation}
We then use the spherical symmetry condition and the fact that
$\bar{\Sigma}_{\mu\nu}$ is time independent outside the horizon to
write the conservation equation (\ref{n1}) in the form

\begin{equation}
\partial_{r^{*}}\bar{\Sigma}_{t}^{r^{*}}+\Gamma^{t}_{tr^{*}}\bar{\Sigma}_{t}^{r^{*}}
-\Gamma^{r^{*}}_{tt}\bar{\Sigma}_{r^{*}}^{t}=0
\label{n3}\end{equation}

\begin{equation}
\partial_{r^{*}}\bar{\Sigma}_{r^{*}}^{r^{*}}+\Gamma^{t}_{tr^{*}}\bar{\Sigma}_{r^{*}}^{r^{*}}
-\Gamma^{t}_{tr^{*}}\bar{\Sigma}_{t}^{t}=0
\label{n4}\end{equation} where
$\bar{\Sigma}_{r^{*}}^{t}=-\bar{\Sigma}_{t}^{r^{*}}$, and
$\bar{\Sigma}_{t}^{t}=\bar{\Sigma}_{\gamma}^{\gamma}-\bar{\Sigma}_{r^{*}}^{r^{*}}$
; $\bar{\Sigma}_{\gamma}^{\gamma}$ is the trace anomaly in
two-dimensions. Using these equations one can write
\begin{equation}\frac{d}{dr}[\bar{\Omega}(r)\bar{\Sigma}^{r^{*}}_{t}] =0
\label{n5}\end{equation}
 and
\begin{equation}
\frac{d}{dr}[\bar{\Omega}(r)\bar{\Sigma}_{r^{*}}^{r^{*}}]=
\frac{1}{2}[\frac{d}{dr}\bar{\Omega}(r)]
\bar{\Sigma}_{\gamma}^{\gamma} \label{n6}\end{equation} the
relation (\ref{n5}) gives immediately
\begin{equation}
\bar{\Sigma}^{r^{*}}_{t}=\alpha \bar{\Omega}^{-1}(r)
\label{n7}\end{equation} with $\alpha$ being an integration
constant. Equation(\ref{n6}) gives

\begin{equation}\label{n8}
\bar{\Sigma}_{r^{*}}^{r^{*}}(r)=( H(r)+\beta )\bar{\Omega}^{-1}(r)
\,\,,\,\,\beta=\bar{\Omega}(L)\bar{\Sigma}_{r^{*}}^{r^{*}}(L)
\end{equation}
where $L=2GM$, and
\begin{equation}\label{x9}
H(r)=\frac{1}{2}\int_{L}^{r}\bar{\Sigma}_{\gamma}^{\gamma}(r')\frac{d}{dr'}\bar{\Omega}(r')
dr' \end{equation} From the equation(\ref{n7})and (\ref{n8}), we
may now write the general form of $\bar{\Sigma}_{\mu}^{\nu}$ as
\begin{equation}
\bar{\Sigma}_{\mu}^{(r)\nu}=\left(\begin{array}{cc}
\bar{\Sigma}_{\gamma}^{\gamma}-\bar{\Omega}^{-1}(r)H(r)  &  0 \\
0  &  \bar{\Omega}^{-1}(r)H(r)
\end{array}\right)+ \bar{\Omega}^{-1}(r)=\left(\begin{array}{cc}
-\beta  &  -\alpha \\
\alpha  &  \beta
\end{array}\right) \end{equation}
If we define $Q=\beta-\alpha$ and $K=\alpha,
\bar{\Sigma}_{\mu}^{\nu} $ takes then the following form

\begin{equation}\label{z1}
 \bar{\Sigma}_{\mu}^{\nu}= \bar{\Sigma}_{\mu}^{(r)\nu}+ \bar{\Sigma}_{\mu}^{(eq)\nu}
\end{equation}
with
\begin{equation}\label{z2} \bar{\Sigma}_{\mu}^{(r)\nu}=\left(\begin{array}{cc}
\bar{\Sigma}_{\gamma}^{\gamma}-\bar{\Omega}^{-1}(r)H(r)  &  0 \\
0  &  \bar{\Omega}^{-1}(r)H(r)
\end{array}  \right)+ K\bar{\Omega}^{-1}(r)
\left(\begin{array}{cc}
-1  &  -1 \\
1 &  1
\end{array}\right)\end{equation}

\begin{equation}\label{w3}
\bar{\Sigma}_{\mu}^{(eq)\nu}= Q\bar{\Omega}^{-1}(r)
\left(\begin{array}{cc}
-1  &  0 \\
0 &  1
\end{array}\right)
\end{equation}

The tensor $\bar{\Sigma}_{\mu}^{(eq)\nu}$ is the traceless part of
$\bar{\Sigma}_{\mu}^{\nu}$ and only $\bar{\Sigma}_{\mu}^{(r)\nu}$
contains off diagonal components.


\end{document}